\documentclass[twocolumn,amsmath,amssymb,superscriptaddress,aps,prl,floatfix,footinbib,longbibliography]{revtex4-2}
\usepackage[]{graphicx}
\usepackage{tabularx}
\usepackage[usenames,dvipsnames]{color}
\usepackage{soul}
\usepackage{bm}

\usepackage{times}
\usepackage{amsfonts}
\usepackage{mathrsfs}
\usepackage{graphicx}% Include figure files
\usepackage{dcolumn}% Align table columns on decimal point
\usepackage{bm}% bold math
\usepackage[dvipsnames]{xcolor}

\usepackage[colorlinks,bookmarks=false,citecolor=blue,linkcolor=red,urlcolor=blue]{hyperref}
\usepackage{multirow}
\usepackage{physics}

\newcommand{\qTF}{q_{\rm TF}}
\begin{document}

%\title{Universal scaling of long-range electronic correlations in Kagome metals}
%\title{Electronic correlations and long-range universal scaling in kagome materials}
%\title{Electronic correlations and long-range universal scaling in kagome metals}
\title{Electronic correlations and universal long-range scaling in kagome metals}

\author{Domenico Di Sante}
\affiliation{Department of Physics and Astronomy, University of Bologna, 40127 Bologna, Italy}\email{domenico.disante@unibo.it}
\affiliation{Center for Computational Quantum Physics, Flatiron Institute, 162 5th Avenue, New York, NY 10010, USA}

\author{Bongjae Kim}
\affiliation{Department of Physics, Kunsan University, Gunsan 54150, Republic of Korea}

\author{Werner Hanke}\affiliation{Institut f\"{u}r Theoretische Physik und Astrophysik and W\"{u}rzburg-Dresden Cluster of Excellence ct.qmat, Universit\"{a}t W\"{u}rzburg, 97074 W\"{u}rzburg, Germany}

\author{Tim Wehling}
\affiliation{Institute of Theoretical Physics, University of Hamburg, Notkestrasse 9, 22607 Hamburg, Germany}
\affiliation{The Hamburg Centre for Ultrafast Imaging, Luruper Chaussee 149, 22761, Hamburg, Germany}

\author{Cesare Franchini}
\affiliation{Department of Physics and Astronomy, University of Bologna, 40127 Bologna, Italy}
\affiliation{Faculty of Physics and Center for Computational Materials Science, University of Vienna, Sensengasse 8, A-1090 Vienna, Austria}

\author{Ronny Thomale}\affiliation{Institut f\"{u}r Theoretische Physik und Astrophysik and W\"{u}rzburg-Dresden Cluster of Excellence ct.qmat, Universit\"{a}t W\"{u}rzburg, 97074 W\"{u}rzburg, Germany}

\author{Giorgio Sangiovanni}\affiliation{Institut f\"{u}r Theoretische Physik und Astrophysik and W\"{u}rzburg-Dresden Cluster of Excellence ct.qmat, Universit\"{a}t W\"{u}rzburg, 97074 W\"{u}rzburg, Germany}

\date{\today}

%\begin{abstract}
%By using the constrained Random Phase Approximation within the framework of first-principles calculations, we investigate the real-space profile of effective Coulomb interactions in a variety of recently synthesized correlated kagome materials, and contrast them to typical correlated perovskite transition metal oxides such as lanthanides. We particularize to KV$_3$Sb$_5$, Co$_3$Sn$_2$S$_2$, FeSn, and Ni$_3$In as representative cases of material classes that show a great variety of correlations-driven fundamental physics. Our selection includes charge-density waves, superconductivity, and magnetism, as well as topologically non-trivial electronic bandstructures. We find that the on-site Coulomb interactions in kagome metals critically depend not only on the screening due to high-energy degrees of freedom, but also on low-energy electronic states that strongly hybridize with the density of states supported by the kagome lattice. Despite then a rather universal real-space decay rescaled by on-site coupling strength across different families of kagome and transition metal oxide perovskite materials, this results in qualitative differences between the long-range interaction profiles of $dp-dp$ and $d-dp$ models. 
%\end{abstract}

\begin{abstract}
We investigate the real-space profile of effective Coulomb interactions in correlated kagome materials. By particularizing to KV$_3$Sb$_5$, Co$_3$Sn$_2$S$_2$, FeSn, and Ni$_3$In, we analyze representative cases that exhibit a large span of correlation-mediated phenomena, and contrast them to prototypical prevoskite transition metal oxides. From our constrained random phase approximation studies we find that the on-site interaction strength in kagome metals not only depends on the screening processes at high energy, but also on the low-energy hybriziation profile of the electronic density of states. Our results indicate that rescaled by the onsite interaction amplitude, all kagome metals exhibit a universal long-range Coulomb behaviour. 
\end{abstract}

\maketitle

\begin{figure}[!t]
\centering
\includegraphics[width=\columnwidth,angle=0,clip=true]{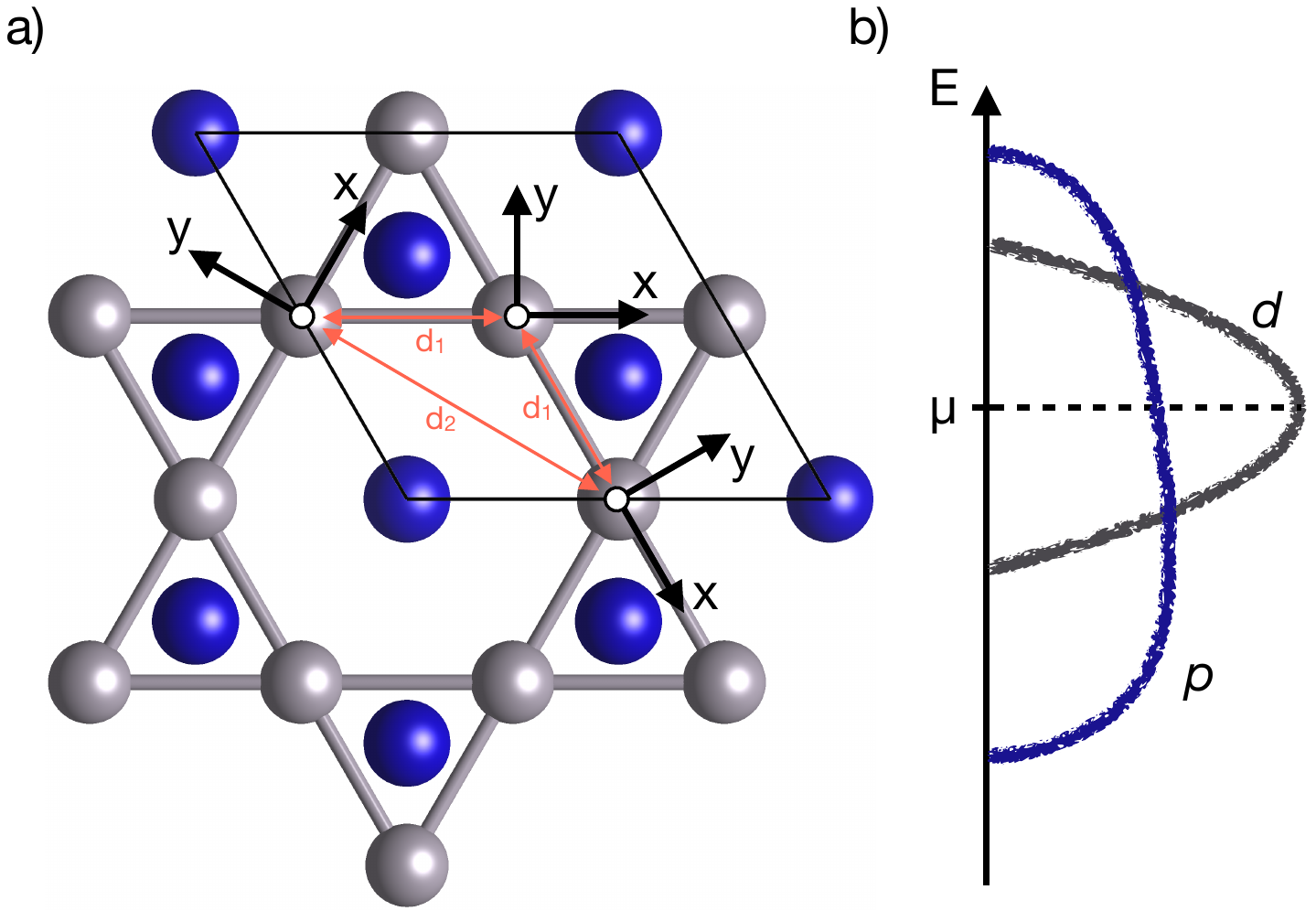}
\caption{a) Top view of a kagome net formed by correlated cation atoms (grey) decorated with non-correlated anion atoms (blue), in the typical arrangement seen in KV$_3$Sb$_5$, Co$_3$Sn$_2$S$_2$, FeSn and Ni$_3$In. Each kagome atom in the unit cell is represented with its respective local $(x,y)-$frame. The red arrows refer to first $(d_1)$ and second $(d_2)$ nearest-neighbour atoms. b) Schematic distribution of the density of states resolved in correlated $(d)$ and non-correlated $(p)$ orbital contributions.}
\label{fig1}
\end{figure}

{\it Introduction.--}The kagome lattice (Fig.~\ref{fig1}a) has emerged as a prototypical playground for sought-after quantum phenomena of electronic matter. From the perspective of localized spins and quantum magnetism, the geometric frustration promoted by the corner-sharing triangles triggers interesting quantum phases such as spin liquids~\cite{NormanRMP2016}. From the viewpoint of bandstructure and itinerant electrons, the kagome lattice offers a whole variety of appealing features at different filling. These range from Dirac cones and van Hove singularities to flat band, where the latter has been suggested as a natural host for ferromagnetism thanks to the quenching of kinetic energy and increase of the density of states~\cite{Mielke1991}, as well as for topological physics~\cite{Parameswaran2013,HuaLeePRB2016,MaPRL2020}. Kagome Dirac cones, on the other hand, have become a paradigmatic tool to accomplish correlated Dirac fermions~\cite{MazinNatComm2014,FuchsJPhysMat2020}, possibly leading to hydrodynamic electron flows meeting the criteria for accessing the turbulent regime~\cite{DiSanteNatComm2020}. Most recently, exotic electron instabilities have been reported at, or close to, van Hove filling, where the nesting properties of the kagome Fermi surface, combined with its sublattice interference, are preeminently suited for enhancing exotic two-particle effective interaction profiles~\cite{KieselPRB2012,KieselPRL2013}.

Despite all these peculiar features tied to the lattice and hence the kinetic theory related to the kagome compounds, the common denominator behind this surprisingly rich and diversified physics and plethora of collective phenomena still is the Coulomb repulsion between electrons. 
%From the spin liquid phase to the unconventional superconductivity, the electronic correlations and, in turn, the balance between kinetic and potential energies are the decisive ingredients for the emergence of a complex phase diagram. 
It is self-evident that reliable first-principles estimations of the Coulomb interaction parameters of kagome materials are of pivotal importance to any attempt to tackle the many-electron problem. Specifically, several avenues of many-body physics born out of a kagome metal parent state are expected to sensitively depend on the precise shape and amplitude of the effective Coulomb interaction profile: For electronically mediated charge density wave formations, the relative strength of the nearest neighbor repulsion appears paramount. Likewise, the harmonic composition of the superconducting pairing function crucially depends on the relevance of long-range interaction components. Finally, while a rescaled fine structure constant concomitant with a preserved algebraic shape of Coulomb repulsion is no prerequisite, it is likely that the true character of long-range Coulomb behaviour has a significant impact on the nature of hydrodynamic electron flow \cite{PhysicsToday2005,PhysicsToday2020}.

In this work, we analyze the Coulomb interaction profile of four representative kagome systems, namely, KV$_3$Sb$_5$, Co$_3$Sn$_2$S$_2$, FeSn and Ni$_3$In. In terms of ground state properties, these prototypical compounds display a large variety of correlations-driven fundamental physics. Specifically, the recently discovered AV$_3$Sb$_5$ (A = K, Rb, Cs)~\cite{OrtizPRM2019} provide an occurrence of kagome metals with the Fermi level tuned to the vicinity of several van Hove singularities with different sublattice character distribution~\cite{WuPRL2021,Kang2022,Hu2022}. When combined with non-local Coulomb correlations, this sublattice decoration gives rise to a profusion of exotic many-body phases, from unconventional superconductivity~\cite{WuPRL2021,KieselPRB2012} to charge (CDW) and spin density-wave (SDW) orders with finite relative angular momentum~\cite{KieselPRL2013}. Indeed, experimental evidence for unconventional manifestations of superconductivity and CDW phases has emerged~\cite{JiangNatMat2021,OrtizPRM2021,LiPRX2021,zhao2021nodal,ChenNat2021}, and any summary on the current status of the field~\cite{NeupertNatPhys2021} conveys a vibrant theoretical and experimental activity devoted to unveiling all correlation-mediated phenomena of AV$_3$Sb$_5$. In turn, the layered half-metal Co$_3$Sn$_2$S$_2$ is a new magnetic Weyl semimetal, where magnetism and topological properties from the Weyl fermions, namely non-trivial Berry curvature and Fermi-arcs, intertwine to create a giant intrinsic anomalous Hall effect~\cite{XuPRB2018,LiuNatPhys2018,WangNatComm2018,LiuScience2019,MoraliScience2019}. FeSn, on the other hand, represents the two-dimensional limit and progenitor of the new family of binary kagome metals T$_m$X$_n$ (T = Fe, Mn, Co ; X = Sn, Ge). These materials are quantum magnets displaying highly-tunable correlated massive Dirac fermions~\cite{YeNat2018,YinNat2018,YeNatComm2019}, and coexistence of Dirac and flat bands as signatures of topology and correlations~\cite{GhimireNatMat2020,KangNatMat2020,LiuNatComm2020}. Finally, Ni$_3$In is characterized by a partially filled flat band that leads to fluctuating magnetic moments with highly unconventional metallic and thermodynamic responses. Moreover, the concomitant observation of non-Fermi liquid behavior also indicates a proximity to quantum criticality~\cite{ye2021flat}.

{\it Methods.--}The common approach to extract interaction parameters from first-principles electronic structure calculations is the constrained Random Phase Approximation (cRPA)~\cite{AryasetiawanPRB2004,MiyakePRB2009,TomczakPRB2009,VaugierPRB2012,Werner2012NatPhys,WernerPRB2015,HansmannPRL2013}. Here, the central quantity is the density susceptibility $\chi(\textbf{r},\textbf{r}',\omega) = \chi^\text{KS}(\textbf{r},\textbf{r}',\omega) + \int\int d\textbf{r}_1d\textbf{r}_2 \chi^\text{KS}(\textbf{r},\textbf{r}_1,\omega) v(\textbf{r}_1,\textbf{r}_2)\chi(\textbf{r}_2,\textbf{r}',\omega)$, where $v(\textbf{r}_1,\textbf{r}_2) = |\textbf{r}_1 - \textbf{r}_2|^{-1}$ is the bare Coulomb potential and
\begin{eqnarray}
\label{eqn1}
\chi^\text{KS}(\textbf{r},\textbf{r}',\omega) &=& \nonumber \\ &\tilde{\sum}_{jj'}&\frac{(f_j - f_{j'}) \psi_{j'}^*(\textbf{r}) \psi_{j'}^*(\textbf{r}') \psi_j(\textbf{r}') \psi_j(\textbf{r})}{\omega - (\epsilon_{j'} - \epsilon_j) + i0^+}
\end{eqnarray}
\noindent is the susceptibility of the non-interacting Kohn-Sham electrons. $\epsilon_j$ and $f_j$ are the energy and occupation of the eigenstate $\psi_j$. $\tilde{\sum}_{jj'}$ excludes from the summation all pairs of states $j,j'$ included in the correlated subspace, defining in our case the $d-dp$ and $dp-dp$ models.
These are distinguished by whether or not the $p$-like states, always included in the construction of the low-energy model, are integrated out in Eq.~\ref{eqn1}. The strength of the interaction is therefore expected to be larger in the $dp-dp$ scheme, as a result of an overall smaller screening.

The screened Coulomb values $W(\textbf{r},\textbf{r}',\omega) = v(\textbf{r},\textbf{r}') + \int\int d\textbf{r}_1d\textbf{r}_2 v(\textbf{r},\textbf{r}_1) \chi(\textbf{r}_1,\textbf{r}_2,\omega) W(\textbf{r}_2,\textbf{r}',\omega)$ determines then the Coulomb tensor as the following static limit
\begin{eqnarray}
\label{eqn2}
U_{\alpha\beta\gamma\delta} &=& \lim_{\omega\rightarrow 0} \int\int d\textbf{r}_1d\textbf{r}_2 w_\alpha^*(\textbf{r}_1) w_\beta^*(\textbf{r}_2) W(\textbf{r}_1,\textbf{r}_2,\omega) \nonumber\\
&\times& w_\gamma(\textbf{r}_1) w_\delta(\textbf{r}_2)
\end{eqnarray}
\noindent where $w(\textbf{r})$ are Wannier functions. We just notice that the restricted summation in Eq.~\ref{eqn1} is straightforwardly implemented only if the target correlated states form an isolated manifold. However, when the correlated states are strongly entangled with the non-correlated ones, as it is the case for the kagome materials under investigation here, $\tilde{\sum}_{jj'}$ must be handled with caution, and specific schemes have been developed~\cite{MiyakePRB2008,MiyakePRB2009,SasiogluPRB2011}.

In this respect, we used the cRPA implementation in VASP~\cite{Kaltak2015,Kresse1996} based on WANNIER90~\cite{Mostofi2008} and the VASP2WANNIER interface~\cite{Franchini2012}. Details of the calculations are listed in Ref.~\cite{comp}, and a dataset of input and output files is publicly released~\cite{zenodo}. Here, we briefly comment on the validity of the cRPA approach, giving a general remark about downfolded low-energy models for electrons, obtained by integrating out the high-energy bands away from the target bands near the Fermi energy. When the gap in these high-energy bands is large, the use of cRPA is justified~\cite{vanLoonPRB2021}.  
In recent works, however, the constrained functional Renormalization Group (cfRG) was studied and contrasted to cRPA~\cite{HonerkampPRB2018,HanPRB2021}. The former goes beyond cRPA by including all one-loop diagrams, which indeed are shown to correct the cRPA tendency towards overscreening, at least for model systems. Nevertheless, it was also shown that cfRG corrections over cRPA remain small when the high-energy bands have a different symmetry than the target bands, and when the number of screening bands is large. This is definitely the case for the kagome metals under investigation here.

\begin{figure}[!t]
\centering
\includegraphics[width=\columnwidth,angle=0,clip=true]{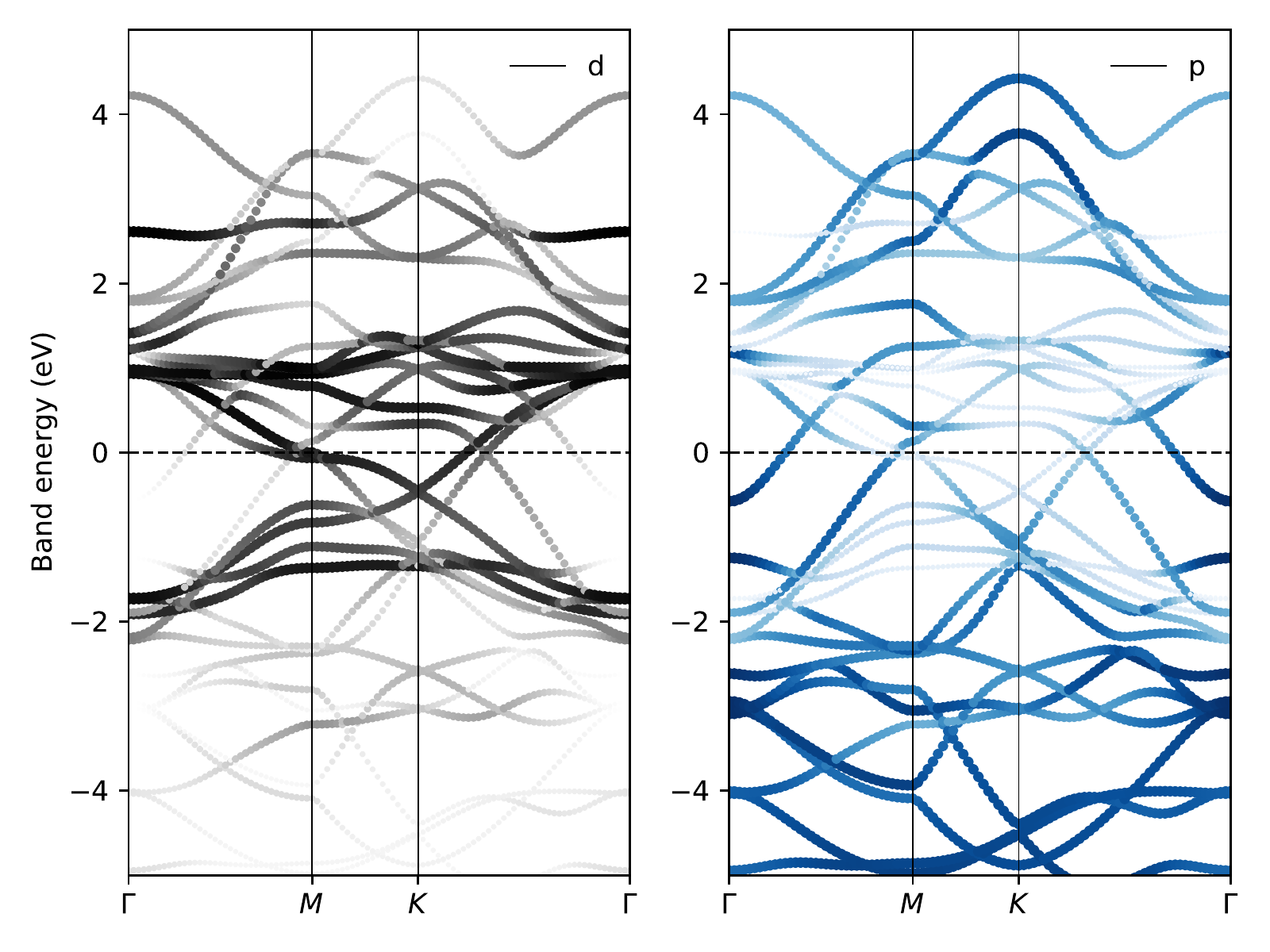}
\caption{Electronic structure of KV$_3$Sb$_5$ along some high-symmetry lines. The left panel highlights the contribution to the Kohn-Sham states from the correlated $d$ orbitals of vanadium atoms, while the right panel highlights the contribution from the non-correlated $p$ states of antimony atoms.}
\label{fig2}
\end{figure}

\begin{table}[!b]
\caption{Averaged values $\bar{\epsilon}_d$ and $\bar{\epsilon}_p$ of the on-site energies for the correlated $d$ orbitals and non-correlated $p$ orbitals with respect to the Fermi energy ($\Delta_{dp} = \bar{\epsilon}_d - \bar{\epsilon}_p$), and diagonally averaged effective on-site Coulomb interactions $U_0 = U(|\textbf{r}| = 0)$ for the $d-dp$ and $dp-dp$ models. All values are given in eV. For the TMO systems, we only considered $t_{2g}$ orbitals as the correlated manifold.}
\begin{tabularx}{\columnwidth}{XXXXXX}
\hline
                  & $\bar{\epsilon}_d$ & $\bar{\epsilon}_p$ & $\Delta_{dp}$ & $U_0^{d-dp}$ & $U_0^{dp-dp}$ \\ \hline
                  &                    &                    &               &              &               \\
%                  &                    &                    &               &              &               \\
KV$_3$Sb$_5$      & 0.2                & -1.2               & 1.4           & 1.7          & 6.3           \\
Co$_3$Sn$_2$S$_2$ & -1.3               & -1.6               & 0.3           & 2.9          & 8.0           \\
FeSn              & -0.5               & 0.1                & -0.6          & 3.0          & 7.6           \\
Ni$_3$In          & -1.2               & 0.2                & -1.4          & 3.3          & 7.2           \\
                  &                    &                    &               &              &               \\
LaTiO$_3$         & 0.5                & -5.6               & 6.1           & 2.2          & 3.3           \\
LaVO$_3$          & 0.2                & -5.4               & 5.6           & 2.4          & 3.7           \\
LaCrO$_3$         & -0.1               & -5.1               & 5.0           & 2.5          & 3.8           \\
%SrVO$_3$          & 0.9                & -4.1               & 5.0           & 3.1          & 6.3           \\
                  &                    &                    &               &              &               \\ \hline
\end{tabularx}
\label{tab2}
\end{table}

{\it Results.--}The table shown in the Supplemental Material~\cite{supplement} reports the Coulomb $U_{\alpha\beta} = U_{\alpha\beta\alpha\beta}$ and exchange $J_{\alpha\beta} = U_{\alpha\beta\beta\alpha}$ interactions in the cubic harmonic basis, where $\alpha$ and $\beta$ refer to the $3d$ orbitals aligned according to the local $(x,y)-$frames of Fig.~\ref{fig1}a). In both models, the Wannier basis includes both $d$ and $p$ states. However, in the $d-dp$ model, the latter are integrated out and actively contribute to the screening of the correlated $d$-manifold. We will see that this aspect plays a crucial role in determining the real-space long-range profile of the interaction.

A feature common to all kagome systems we have investigated is the high degree of covalent bonding, manifested by the hybridization between correlated $d$ states and non-correlated $p$ orbitals, as schematically shown in Fig.~\ref{fig1}b. A more quantitative analysis, for the case of KV$_3$Sb$_5$, is shown in Fig.~\ref{fig2}, but it is likewise valid for all kagome materials here. On both sides of the chemical potential (reference zero energy in Fig.~\ref{fig2}), the electronic states deriving from the $d$ orbitals of vanadium atoms are sandwiched between the antimony $p$-states. According to the energy difference in the denominator of the density susceptibility of Eq.~\ref{eqn1}, which enters the system polarizability in the definition of the cRPA dielectric constant, this gives rise to a strong reduction of the on-site Coulomb interaction $U_{\alpha\beta}$ in the $d-dp$ model as a result of an enhanced screening compared to the $dp-dp$ model. However, only the on-site Coulomb matrix is affected by such reduction. The values of the Hund's couplings $J_{\alpha\beta}$
are, to a good extent, insensitive to the model.

Even more interestingly, a marked difference between the electronic properties of kagome systems and typical transition metal oxides (TMO), such as for instance lanthanide oxides~\cite{KimPRB2018}, is the presence of $p$-like bands crossing the Fermi level. For KV$_3$Sb$_5$, the antimony $p_z$ orbitals form an electron pocket at the $\Gamma$ point and sizeably contribute to the Dirac bands along the $\Gamma M$ and $\Gamma K$ lines. The differences between the kagome systems and typical TMO materials can be made on a quantitative ground by estimating the energy gap $\Delta_{dp} = \bar{\epsilon}_d - \bar{\epsilon}_p$ separating the averaged values of the $d$ and $p$ orbitals local levels $\bar{\epsilon}_d$ and $\bar{\epsilon}_p$ (not to be confused with the charge transfer energy of superconducting TMO as cuprates and nickelates, defined as the energy to promote an electron from the ligand states to the correlated $d$ manifold, {\it i.e.,} $d^9 \rightarrow d^{10}${\underline{L}}). These numbers are expected to reflect the energy distribution of the orbital resolved density of states $g_\alpha(\epsilon)$, given by the band centroid $\int d\epsilon \epsilon g_\alpha(\epsilon) / \int d\epsilon g_\alpha(\epsilon)$ for orbital $\alpha = d,p$.

In typical TMO, the value of $\Delta_{dp}$ is positive and amounts to several eV. As shown in Table~\ref{tab2}, on the other hand, the kagome systems are characterized by a much smaller $d-p$ gap which indeed reflects the higher degree of covalency with respect to TMO. Moreover, for FeSn and Ni$_3$In, $\Delta_{dp}$ is negative, suggesting that on average the on-site energies of the non-correlated orbitals are higher than the ones of the correlated manifold. A sensibly smaller $\Delta_{dp}$ for the family of kagome materials determines then a ratio $U_0^{dp-dp}/U_0^{d-dp}$ which exceeds by more than a factor of two that for TMO.   

Beyond the local interaction, the kagome lattice, with three atomic sites per unit cell, offers itself as a perfect setting to compute the real-space profile of the Coulomb repulsion. According to the centers of Wannier functions entering Eq.~\ref{eqn2}, besides the on-site Coulomb parameters $U_{\alpha\beta}(|\textbf{r}|=0)$ given in Ref.~\cite{supplement}, we can easily compute the first nearest-neighbour $U_{\alpha\beta}(|\textbf{r}|=d_1)$ and second nearest-neighbour $U_{\alpha\beta}(|\textbf{r}|=d_2)$ parameters, where $d_1$ and $d_2$ are drawn in Fig.~\ref{fig1}a). In the case of Ni$_3$In, the peculiar bond-length modulation of the triangles allows for the estimate of two, quite close, first nearest-neighbour distances~\cite{ye2021flat}.

\begin{figure}[!t]
\centering
\includegraphics[width=\columnwidth,angle=0,clip=true]{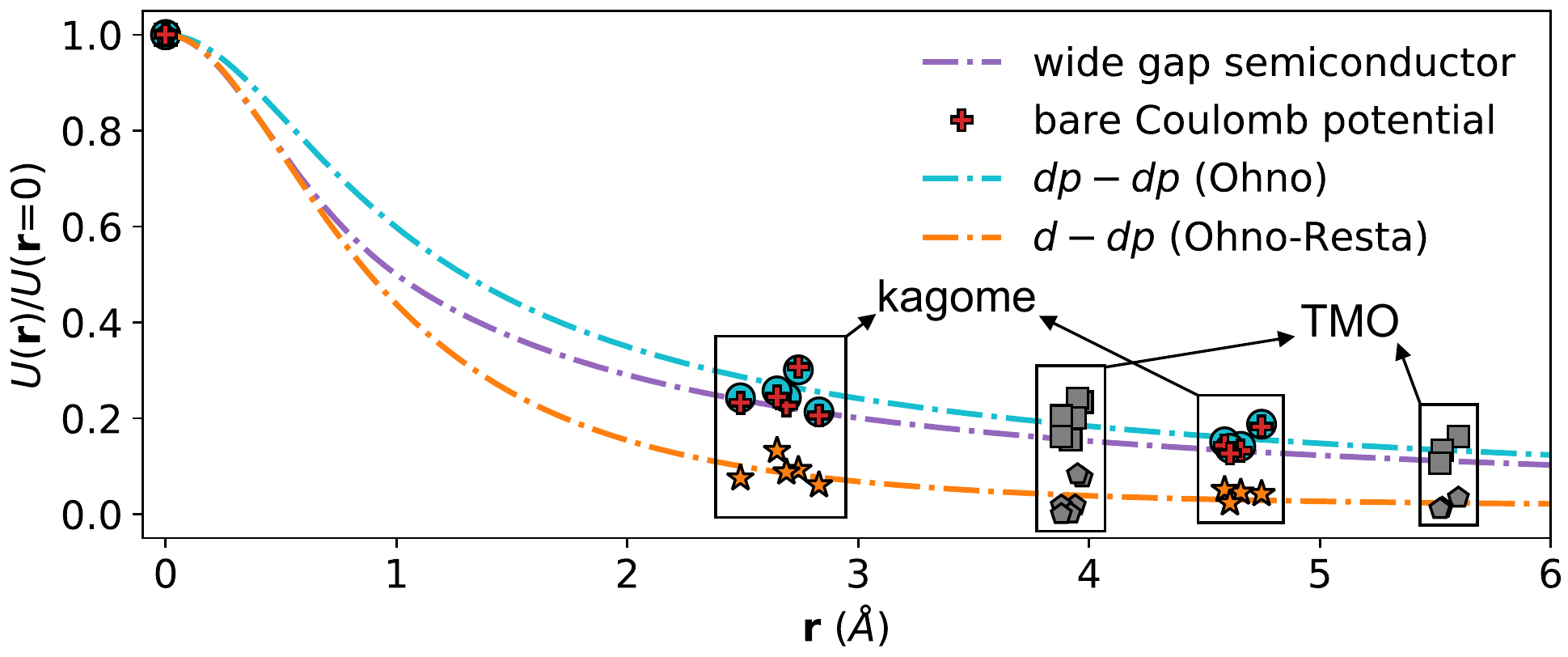}
\caption{Real-space decay of the Coulomb interaction in the kagome systems for the bare potential (red crosses), $dp-dp$ (blue circles) and $d-dp$ (orange stars) models, as well as in LaTiO$_3$, LaVO$_3$ and LaCrO$_3$ (grey squares for $dp-dp$ and pentagons for $d-dp$ models, respectively). The dotted-dashed lines are fits according to the screening models described in the main text.}
\label{fig3}
\end{figure}

In Fig.~\ref{fig3} we show, for both models, the real-space profile of the long-range Coulomb interaction, where for the sake of clarity, we have averaged the diagonal of the Coulomb tensor according to $U(\textbf{r}) = (1/5)\sum_{\alpha=1}^5 U_{\alpha\alpha}(\textbf{r})$. 
We observe that, upon rescaling by the corresponding on-site strengths, the two models display a rather universal real-space decay among the different families of kagome materials, and even the TMO. This universal decay of the rescaled Coulomb interactions depends only on the constraining scheme used in the RPA density susceptibility $\chi(\textbf{r},\textbf{r}',\omega)$ derived from Eq.~\ref{eqn1}, which can be understood as follows. According to Eq.~\ref{eqn2}, $U(\textbf{r})$ is determined by $d$-electron Wannier functions and the (partially) screened Coulomb interaction $W(\textbf{r},\textbf{r}',\omega)$. For simplicity we consider the static limit and assume $W(\textbf{r},\textbf{r}',\omega=0)=W(|\textbf{r}-\textbf{r}'|)$. For a typical spatial extent $\delta$ of the Wannier functions, we estimate the averaged Coulomb interaction $U(r)=U(|\textbf{r}|)\approx W(\sqrt{r^2+\delta^2})$.

If the screening is local, i.e. $W(r)=\frac{e^2}{\epsilon_1 r}$ with dielectric constant $\epsilon_1$, one can show that $U(r)=\frac{e^2}{\epsilon_1 \sqrt{r^2+\delta^2}}$, which is also called Ohno potential \cite{OhnoTCA1964, DewarTCA1977} and leads to 
\begin{equation}
U(r)/U(r=0)=\frac{1}{\sqrt{r^2/\delta^2+1}}.\label{eqnOhno}
\end{equation}
Fig.~\ref{fig3} shows that the Ohno potential alone describes the real-space profile of the bare Coulomb interaction and in the $dp-dp$ very well, using $\delta\approx 0.75$ \AA.

The decay of the Coulomb interaction with the distance in the $d-dp$ models falls instead on another universal curve, which is indicative of non-local screening effects. These can be understood in the Resta model \cite{RestaPRB1977}, which is a generalization of Thomas-Fermi theory to describe screening from gapped to metallic systems. The central idea of the Resta model is that in gapped systems screening charges can still be moved but only up to a certain ionic radius $R$. For metals we have $R\rightarrow\infty$ and $R$ decreases with increasing gap size. In terms of the Thomas-Fermi wave number $\qTF$, the ``Resta screened" interaction reads $W(r)=\frac{e^2}{\bar\epsilon(r) r}$, where $$\bar\epsilon(r)=\epsilon_1\cdot\left\{\begin{matrix}
\qTF R/[\sinh(\qTF (R-r))+\qTF r], & r\leq R\\
1, & r\geq R
\end{matrix}\right.$$
and $\epsilon_1=\sinh(\qTF R)/\qTF R$. Beyond the ionic radius $R$, the screening appears local, with finite dielectric constant $\epsilon_1$. If the Thomas-Fermi screening length $1/\qTF$ is substantially shorter than the ionic radius, we recover the usual Thomas-Fermi screening, i.e. $\bar\epsilon(r)\approx e^{\qTF r}$, for $r\ll R$.

This analysis of the cRPA data allows us to assess that the screening in the $dp-dp$ models resembles that of a wide-gap semiconductor, which leads to ionic radii $R\lesssim 2.5$ \AA~\cite{RestaPRB1977}. This is why the $dp-dp$ models are well captured both by a model of purely local screening or more generally by a type of screening as the one found in wide gap semiconductors (e.g. $R=1.5$ \AA\, and $\qTF=2.0$ \AA$^{-1}$).
In the $d-dp$ models instead, the cRPA screening mimics systems with smaller gap or even metallic systems and correspondingly larger ionic radii. Indeed, we can describe the real space profile of Coulomb interaction in the $d-dp$ models, e.g. with $R=6.0$ \AA\, and $\qTF=0.75$ \AA$^{-1}$. While other choices of $R$ and $\qTF$ can also describe the $d-dp$ models, all of these combinations have in common that ionic radii always exceed the interatomic distances making the screening appear non-local and (almost) metallic at (next) nearest-neighbor distances.

{\it Outlook.--}Our work represents a coherent attempt of giving a broad overview on the Coulomb interaction parameters for several different kagome systems, which have been experimentally realized and characterized only very recently. The realistic description of these kagome metals, at odds with simplified modelings, underlines the importance of their multi-orbital nature. Different groups of correlated orbitals give origin to band structure features of different bandwidths and energy alignments, with a strong orbital-selectivity~\cite{LiuNatComm2020}. First steps to shed light onto the possible role of correlation in these new materials based on DFT+DMFT have been recently reported~\cite{LiuNatComm2020,HuangPRB2020,LiNatComm2021,ZhaoPRB2021,Xie2021}. There, a strong assumption is made in favour of a local character of the electron-electron interaction. In addition, within such a local assumption, only simplified Coulomb tensors have been used so far, in terms of averaged values $U$ and $J$ only. Nonetheless, an interesting correlated multi-orbital physics has already been put forward. Besides the orbital selective Dirac fermions, there are evidences that Hund’s rule coupling can be effective in favoring high-spin configurations, even though the high degree of covalency makes an atomic-like description of the physics tricky. 
%\cite{ZhaoPRB2021,WernerPRL2008}. 
Current numerical capabilities of DFT+DMFT implementations allow for treating the full structure of the Coulomb repulsion $U_{\alpha\beta\gamma\delta}$. The inclusion of the non-local interaction terms calculated here, poses instead some conceptual as well as technical challenges. Yet, these need to be addressed, as the Coulomb terms beyond the local one, can by no means be neglected in these classes of metals. The kagome materials, with their pronounced orbital-selectivity, will become paradigmatic settings for such a level of sophistication.

From the viewpoint of non-local correlations, known already to be crucial in the context of other frustrated geometries, such as the triangular lattice~\cite{HansmannPRL2013,CaoPRB2018}, the model-dependent real-space scaling that we have derived from our cRPA calculations will help in pinpointing the many-body instabilities expected for the kagome Hubbard model (KHM). Here, the sublattice interference mechanism plays a crucial role in suppressing the strength of the local interaction~\cite{KieselPRB2012}. Indeed, the phase diagram of the $U_0 - U_1$ KHM is decisively determined by the ratio between local $U_0$ and first nearest-neighbor $U_1$ contributions, with several superconducting particle-particle and unconventional particle-hole instabilities dominating wider regions of the phase space~\cite{KieselPRL2013}. Beyond the idealized one-orbital kagome model, recent RPA many-body calculations have shown, for a more realistic model of the AV$_3$Sb$_5$ kagome systems, that not only $U_0$ but also the $U_1/U_0$ ratio guides a crossover between several unconventional pairing symmetries of the superconducting order parameter, from nodeless $d+id-$ to nodal $f-$waves~\cite{WuPRL2021}.

In conclusion, our understanding of the real-space profile of the screened Coulomb interaction is expected to give fundamental guidance to future investigations of the role of electronic correlations in kagome metals, a class of materials that has the potential to rewrite the paradigms of electronically mediated many-body instabilities.

{\it Acknowledgments.--}The authors are grateful to Jan Tomczak and Philipp Hansmann for invaluable comments and suggestions. The research leading to these results has received funding from the European Union’s Horizon 2020 research and innovation programme under the Marie Sk{\l}odowska-Curie Grant Agreement No. 897276. This work is funded by the Deutsche Forschungsgemeinschaft (DFG, German Research Foundation) through Project-ID 258499086 - SFB 1170, through the W{\"u}rzburg-Dresden Cluster of Excellence on Complexity and Topology in Quantum Matter-ct.qmat Project-ID 390858490 - EXC 2147 as well as the Cluster of Excellence ‘CUI: Advanced Imaging of Matter' – EXC 2056 (Project No. 390715994). 
B.K. acknowledges support by NRF grant (No. 2021R1C1C1007017 and No. 2022M3H4A1A04074153) and KISTI supercomputing Center
(Project No. KSC-2021-CRE-0605).
The authors gratefully acknowledge the Gauss Centre for Supercomputing e.V. for funding this project by providing computing time on the GCS Supercomputer SuperMUC at Leibniz Supercomputing Centre. Supercomputing time on the Vienna Scientific cluster (VSC) is also gratefully acknowledged. The Flatiron Institute is a division of the Simons Foundation.

\bibliography{biblio}

\begin{thebibliography}{64}
\bibitem [{com()}]{comp}%
  \BibitemOpen
  \href@noop {} {}\bibinfo {note} {We included around 200 unoccupied bands, a
  $9\times9\times6$ $\bm{k}$-mesh ($9\times9\times9$ only for Co$_3$Sn$_2$S$_2$
  due to its rhombohedral lattice structure with the space group $R\bar{3}m$)
  and a 600 eV cutoff to calculate the dielectric function and properly include
  local-field effects. The Perdew-Burke-Ernzerhof functional revised for solid
  is used in all calculations~\cite{PBEsol}. Partial occupancies $f_j$ in
  Eq.~\ref{eqn1} are computed by the method of Methfessel-Paxton with a
  smearing of 0.2 eV. In the construction of Wannier functions, $d$ and $p$
  orbitals are always considered, and the following energy windows for the
  disentanglement are set: $[-5.5,5.0]$ eV for KV$_3$Sb$_5$, $[-8.2,7.8]$ eV
  for Co$_3$Sn$_2$S$_2$, $[-5.7,6.8]$ eV for FeSn, and $[-4.9,7.6]$ eV for
  Ni$_3$In. No optimization of Wannier functions spread is performed after the
  disentanglement.}\BibitemShut {Stop}%
\end{thebibliography}

\end{document}